\documentclass{cs20proc}

\usepackage{kantlipsum}

\editors{S. J. Wolk}
\publisher{Zenodo}
\conference{The 20th Cambridge Workshop on Cool Stars, Stellar Systems, and the Sun}
\conferencedate{2018}

\title{The Evolution of $T = 10^4$ K Blackbody-Like Continuum Radiation in the Impulsive Phase of dMe Flares}
\author{Adam Kowalski,$^{1,2, 3}$, Mihalis Mathioudakis$^{4}$, Suzanne L. Hawley$^{5}$}

\affiliation{$^{1}$ Department of Astrophysical and Planetary Sciences, University of Colorado Boulder, 2000 Colorado Ave, Boulder, CO 80305, USA. \\
			 $^{2}$ National Solar Observatory, University of Colorado Boulder, 3665 Discovery Drive, Boulder, CO 80303, USA. \\
			 $^{3}$ Laboratory for Atmospheric and Space Physics, University of Colorado Boulder, 3665 Discovery Drive, Boulder, CO 80303, USA. \\
			 $^{4}$Astrophysics Research Centre, School of Mathematics and Physics, Queen's University, Belfast, BT7 1NN, Northern Ireland, U.K.\\ 
			 
$^{5}$ University of Washington Department of Astronomy, 3910 15th Ave NE, Seattle, WA 98195, USA.}
\onecolumn

\shorttitle{Optical Color Temperature in dMe Flares}
\shortauthors{Kowalski, et al.\ }

\abs{The near-ultraviolet and optical (white-light) continuum radiation in M dwarf flares exhibits a range of observed characteristics, suggesting that the amount of heating at large optical depth varies among impulsive-type and gradual-type flares.  Specific flux ratios from high-time cadence spectra and narrowband continuum photometry have also shown that these characteristics vary from the peak to the gradual decay phases of flares.  In these proceedings, we present the highest-time cadence ($\sim1$~s), highest signal-to-noise ($\sim100$) constraints on the optical color temperature evolution during the rise phase of a large, impulsive-type dMe flare event.  The flare exhibits compelling evidence of a hot, color temperature ($T \sim 10,000$ K), but the Balmer jump ratios show that the flare cannot be explained by isothermal slabs or blackbody surfaces at any time in the evolution.   The new data analysis establish these properties as critical challenges for any flare model, and we discuss 1D radiative-hydrodynamic modeling that will be compared to the evolution of the flare colors in this intriguing event.    }

\begin{document}

\maketitle

\section{Introduction}
Stellar flares are an important probe of the physics of explosive phenomena in the atmospheres of stars.  Magnetically active M dwarf (dMe; diagnosed by H$\alpha$ in emission in quiescence) have steady-state X-ray luminosities that are 20x larger than the Sun's at solar maximum \citep[e.g.,][]{Ayres2015B, Robrade2005, MitraKraev2005}, and they occasionally produce flares that are $100-1000$x more energetic than the largest flares on the Sun \citep{Kowalski2010, Osten2010, Osten2016}.  Due to their small sizes and cool photospheric temperatures, mid-M dwarfs, such as the dM4.5e star YZ CMi, are targeted to study the physics of ultraviolet, blue, and optical flare radiation.

There are many properties of  flare spectra that are not yet understood in terms of atmospheric heating and dynamic processes.  The optical continuum spectra in the impulsive phase of some dMe flares exhibit a color temperature of a $T\sim10^4$ K blackbody \citep{HP91, Kowalski2013, Kowalski2016}.  This phenomenon is well observed during large flares \citep[see also][]{Fuhrmeister2008}, but it also appears in some lower amplitude, lower energy events \citep{Kowalski2016} that are comparable to the energies of large solar flares.  This white-light property is a critical test for any flare model, since the continuum radiation originates in the deepest and densest layers that are heated significantly in the flaring atmosphere. The true nature of this white-light continuum component can only be understood in terms of formation in an actual, dynamic stellar flaring atmosphere, not from uniform slabs.

The pan-chromatic properties of the $T=10^4$ K blackbody-like radiation at FUV and NUV wavelengths are important for modeling ozone photochemistry in habitable zone planets around M dwarfs \citep{Segura2010, Ranjan2017, Howard2018, Tilley2017, Loyd2018A}; extrapolating to IR bands \citep{Davenport2012} is important for predicting what JWST will serendipitously observe during transit spectroscopy.  
Radiative-hydrodynamic (RHD) flare models that have been constrained by observations can be used to extrapolate to wavelength regimes that are difficult to observe regularly.  
At Cool Stars 20,  we presented high-time cadence flare continuum photometry and
comparisons to a new grid of dMe flare models with very high electron beam energies.   In these
proceedings to the workshop, we present the highest time-cadence
constraints yet on the formation and persistence of
hot, $10^4$ K blackbody-like continuum radiation in the rise
phase of a large dMe flare; the grid of models will be presented in Kowalski et al. 2018, \emph{in prep.}, and will be available online as for the F-CHROMA grid of solar flare RHD models\footnote{\url{https://star.pst.qub.ac.uk/wiki/doku.php/public/solarmodels/start}.}.


\section{Observations}
The peak flare data of 20 dMe flares observed through three custom narrowband continuum filters were analyzed in \citet{Kowalski2016}.  These data were obtained with the frame-transfer instrument ULTRACAM \citep{Dhillon2007} on the William Herschel and New Technology Telescopes. Very high-time cadence ($\sim 0.1-2$~s) and high signal-to-noise light curves were presented.  The specific flare-only flux densities ($F_{\lambda}^{'}$) were 
calculated in each filter and were used to construct flare color measurements for direct comparison to radiative-hydrodynamic model atmospheres, blackbody surfaces, and uniform slab models.   The Balmer jump ratio (FcolorB) was calculated as the ratio of specific flux in the NBF3500 filter to the NBF4170 filter.  The optical color temperature was calculated from the ratio (FcolorR) of the specific flux in the NBF4170 filter to the RC\#1 ($\lambda_{c} = 6010$ \AA) filter, giving $T_{\rm{FcolorR}}$.  
However, the high-time resolution colors have not yet been analyzed in detail for the 
 largest two events in the sample:  the IF1 and IF3 events on YZ CMi.  Here, we analyze the time-evolution of $T_{\rm{FcolorR}}$ for the IF1 event, which produced a peak flux enhancement of $\sim31$ in NBF3500 (slightly bluer than the central wavelength of the broader Johnson $U$-band) at 22:33:54 UT on 2012 Jan 13.  The $U$-band peak luminosity for this event is estimated to be $\approx10^{30}$ erg s$^{-1}$.

The NBF4170 flux enhancement evolution of the IF1 event at 1.1~s time-cadence is shown in Figure \ref{fig:yzcmi}(a).  In the rise phase, there are three distinct episodes of flux enhancement.  The slowest rise phase lasts 30~s, the medium-fast rise lasts 10~s, and the very fast rise (hereafter, ``fast rise'') lasts 20~s.  A lower amplitude, longer-duration event is evident as a ``secondary flare'' in the decay phase.  A much larger energy, higher amplitude event (IF3 in \citet{Kowalski2016}) occurs 10 minutes later with a peak flux enhancement of $\sim30$ in NBF4170.  The detailed analysis of the colors in the secondary flare and the larger event will be presented elsewhere.

\section{Analysis} \label{sec:analysis}
\begin{figure}[ht]

	\includegraphics[width=0.48\linewidth]{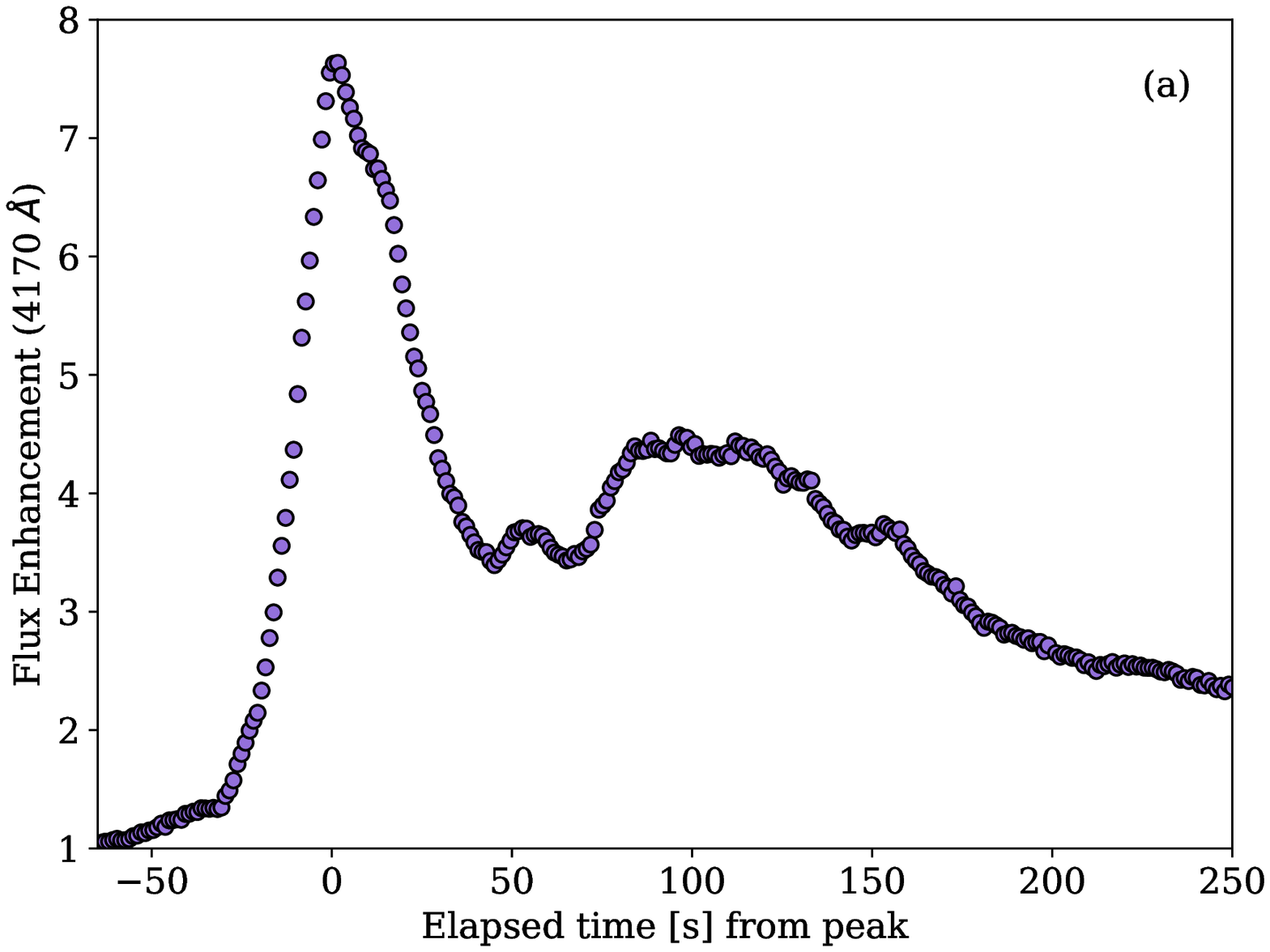}
	\includegraphics[width=0.48\linewidth]{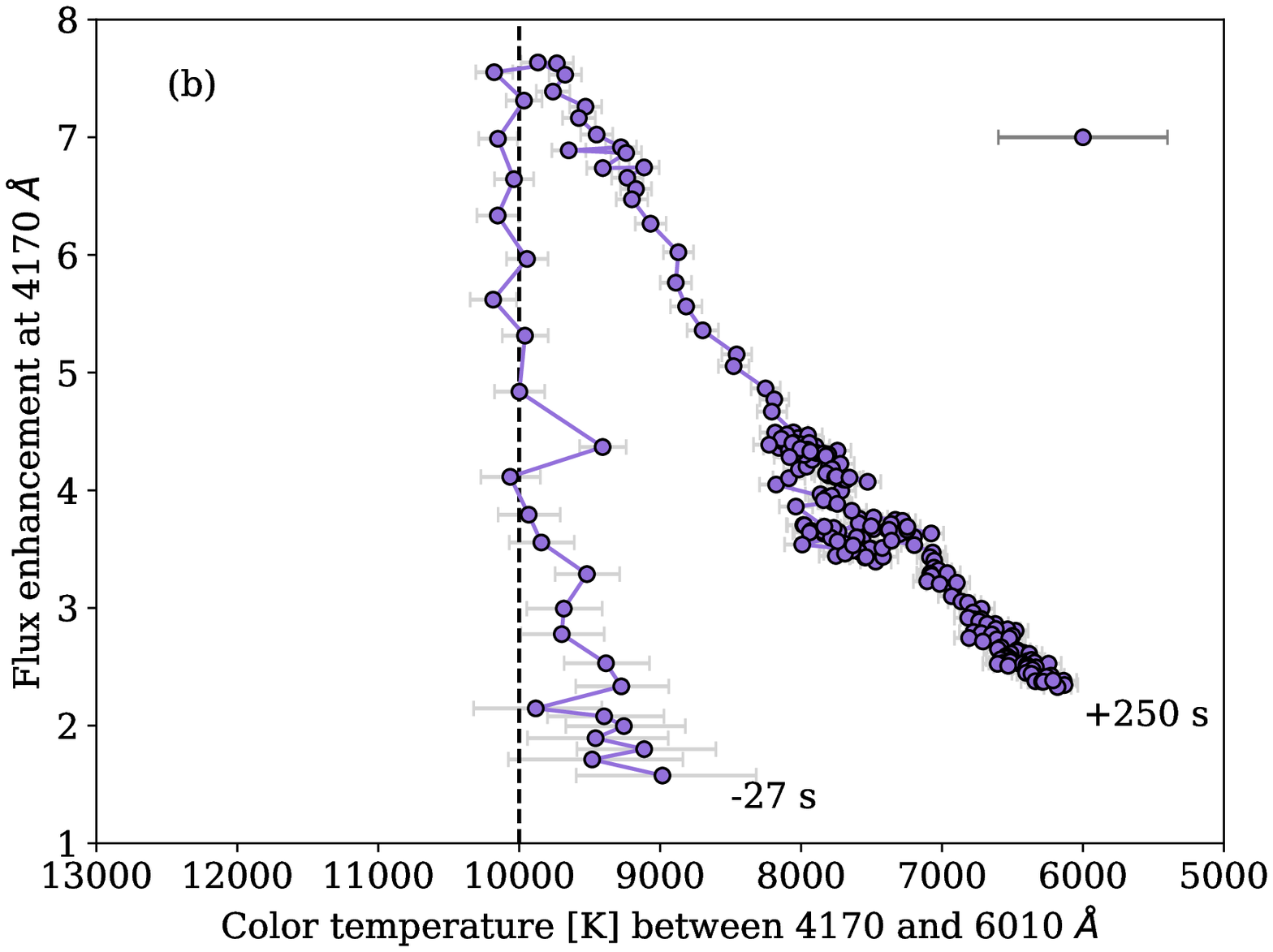}
	\caption{\textbf{(a)} ULTRACAM data of the blue ($\lambda = 4145 - 4195$ \AA) continuum flux enhancement during a large flare on YZ CMi from \citet{Kowalski2016}.  The cadence is 1.1~s, and $t=0$~s corresponds to 22:33:54 UT on 2012 Jan 13.  \textbf{(b)} Flux enhancement in the blue flare continuum vs.\ the color temperature ($T_{\rm{FcolorR}}$) fit to the ULTRACAM data.  An optical color temperature of $\ge10,000$ K is clearly and unambiguously attained in the middle of the fast rise phase.  These values of color temperature may be lower estimates due to the so-called ``conundruum'' radiation \citep[see the IF3 event in][]{Kowalski2013}.  A representative 1$\sigma$ systematic uncertainty in the absolute flux calibration in the rise phase is shown in upper right, while the error bars on the data are the statistical errors only.   This display of flux vs.\ temperature is inspired by \citet{Thomson2005}, who showed the evolution of solar flare GOES flux vs. coronal temperature in a similar way.  In contrast to the optical continuum color temperature evolution of this dMe flare, the maximum coronal temperatures in solar flares occur in the mid-rise phase of the GOES flux enhancement and decrease by the peak (which is qualitatively in line with the Neupert effect).  }
	\label{fig:yzcmi}
\end{figure}

To parameterize the evolution of the flare color temperature, we fit a simple blackbody to the $4170-6010$ \AA\ continuum color (FcolorR; $F_{\rm{4170}}^{'} / F_{\rm{6010}}^{'}$).  This gives an optical color temperature $T_{\rm{FcolorR}}$.   Figure \ref{fig:yzcmi}(b) shows the highest-ever time-resolution, highest-ever signal-to-noise constraints on the development and persistence of the $10^{4}$ K blackbody-like radiation in the rise phase of a large dMe flare\footnote{See \citet{Zhilyaev2007} for very high-time resolution ``colorimetry'' during moderate amplitude flares; these data were obtained in broadband filters, which are difficult to interpret \citep{Allred2006}.}.  From the blue optical to the red optical wavelength regimes, the color temperature increases above $T=8500$ K in the early rise phase when the flux enhancement exceeds a factor of two.  By the mid-rise phase, the color temperature is very clearly at $T \sim 10,000$ K, and it sustains at this value for each 1.1~s exposure time through the peak.  The light curve of the 4170 \AA\ continuum flux (Figure \ref{fig:yzcmi}(a)) shows no sub-structure in the fast rising period at this time-resolution.  This is the first time that light curve evolution and color temperature have been mapped in this detail in the rise phase.

The high-time resolution also allows us to determine that the  color temperature falls  during the peak and at the onset of the fast decay phase.  At similar flux levels, the color temperature is significantly lower in the fast decay phase than in the fast rise phase.  This evolution of the color temperature has been discussed for a larger event\footnote{The IF3 event in \citet{Kowalski2013}.} in \citet{Kowalski2013} and a smaller event\footnote{The IF4/F2 event in \citet{Kowalski2016}.} with ULTRACAM photometry in \citet{Kowalski2016}, both on YZ CMi.  In the IF1 event here, we have characterized this impulsive-phase evolution at 15x higher time cadence (1.1~s) than done before with low-resolution spectra and 10x higher signal-to-noise than a smaller event already discussed in \citet{Kowalski2016}.  After the fast decay phase ends, a secondary flare occurs from $t\sim50$~s to $t\sim180$~s with evident sub-structure.  

Because this event is so large, the statistical errors in the colors are close to 1\% near the peak.  
A representative systematic uncertainty of 5\% in the color temperature in the rise phase is shown as an error bar in upper right of Figure \ref{fig:yzcmi}(b).  Accounting for a conservative 2$\sigma$ error in the systematic flux calibration, the IF1 event attains an optical color temperature of at least 8800 K in the rise phase. 
 There are other systematic uncertainties due to fitting the color temperature over the large wavelength range of $4170-6010$ \AA\ \citep{Kowalski2013,Kowalski2016}, such as the possible presence of Balmer line wings in the blue and a cooler continuum component in the red \citep{Kowalski2013},
which together make the color temperature values $T_{\rm{FcolorR}}$ in Figure \ref{fig:yzcmi}(b) likely lower limits to the actual blue-optical, blackbody color temperature, $T_{\rm{BB}}$, by at least 1000 K.  Even considering these possible sources of systematic uncertainty, the data shown in Figure \ref{fig:yzcmi} exhibit the strongest evidence to date for the presence of $T \sim 10^4$ K blackbody-like radiation at very high time cadence in dMe flares.

\begin{figure}[ht]
\centering
	\includegraphics[width=0.9\linewidth]{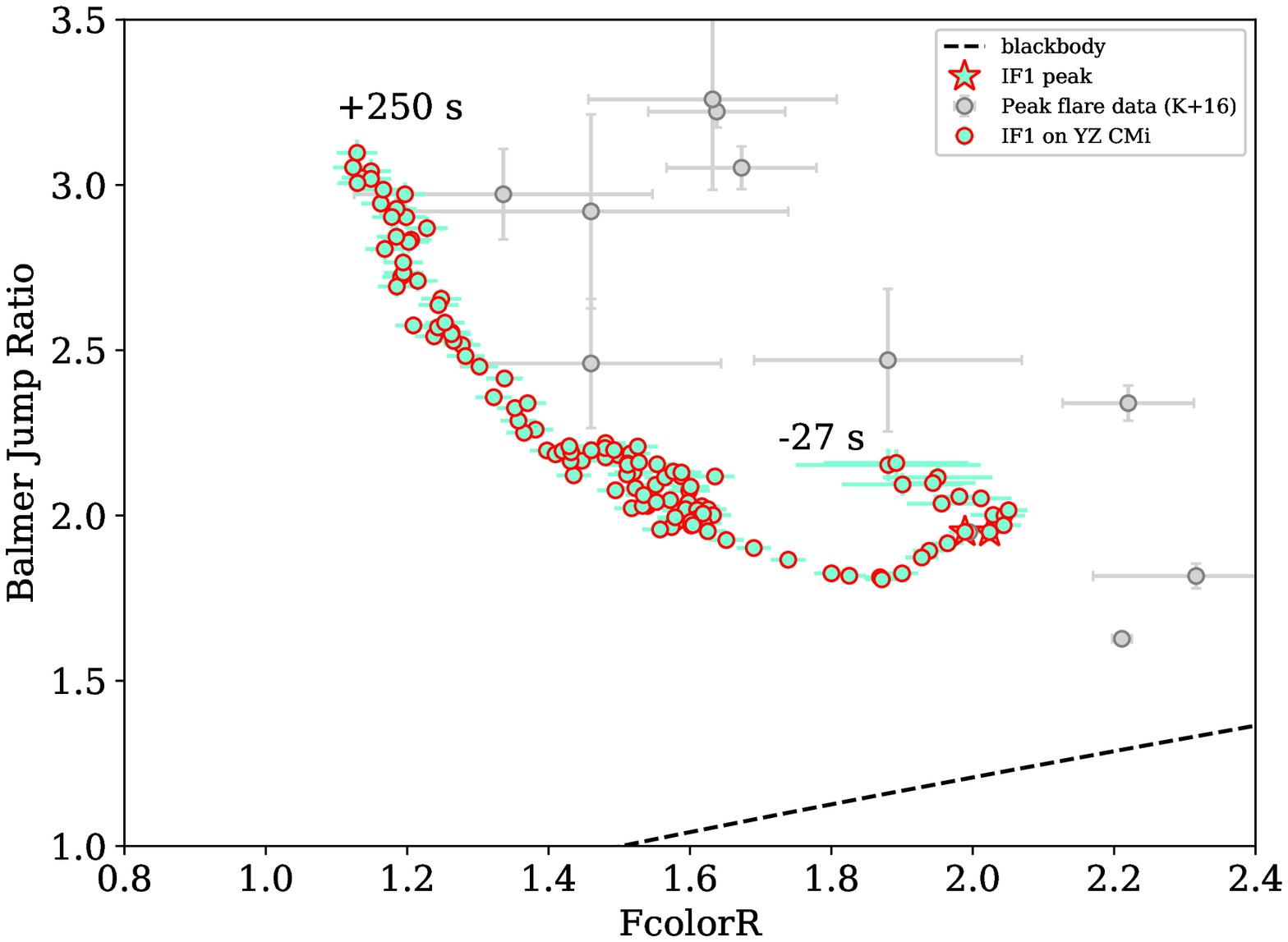}
	\caption{Flare color-color diagram inspired by the broadband color-color diagrams of \citet{Zhilyaev2007}. This figure shows the Balmer jump ratio vs.\ the blue-to-red continuum color for the peaks of the dMe flare sample in \citet{Kowalski2016} (\emph{gray}).  The time-evolution of the IF1 flare colors at a cadence of 2.2~s is shown as the red circled outlines.  A clockwise evolution occurs in the IF1 event.  For similar values of the color temperature, the Balmer jumps are smaller in the fast decay phase than in the rise phase.  A rapid decrease in FcolorR is a characteristic change that occurs during the end of the peak phase (peak phase denoted by star symbols) and the start of the fast decay phase.  Note, the error bars on all flare colors shown here are the statistical errors (not the systematic errors in the online data at Zenodo), which show the significant intraflare variations.  }
	\label{fig:yzcmi_colors}
\end{figure}

  Figure \ref{fig:yzcmi_colors} is a color-color diagram that shows the evolution of the Balmer jump ratio and FcolorR (blackbody-like color temperature).  The colors corresponding to blackbody radiation are indicated as the dashed line, while the Balmer jump ratios from uniform $T\le20,000$ K hydrogen slabs with low optical depths are at significantly larger values than the $y$-extent of the figure \citep[e.g.,][]{Kunkel1970}.  Clearly, neither optically thin slabs nor blackbody surfaces explain the colors of IF1.  Balmer and Pachen opacities modify the emergent radiation from optically thin continuum emission, while variations in temperature and electron density over atmospheric depth modify the properties of the emergent radiation from a blackbody surface \citep{Kowalski2015, Kowalski2015IAU}.  The colors are somewhat closer to  blackbodies of $T\sim10,000$ K, and thus we describe the optical continuum as a ``hot, blackbody-like" continuum.   
  
  For similar color temperature values ($T_{FcolorR}$) in the rise phase and fast decay phases, the Balmer jump ratio is smaller (closer to the blackbody curve) in the fast decay phase.  There is a change in slope in the 4170 \AA\ flux enhancement in the fast decay phase, which is not seen as prominently at the other continuum wavelengths (see discussion of this feature in \citet{Kowalski2016}). The minimum Balmer jump ratio occurs during this light curve feature.
  
  The evolution of the flare radiation in this color-color space at such high-time resolution provides significant insight into modeling the actual time evolution of flare light curves and colors using radiative hydrodynamic (RHD) models that incorporate atmospheric physics that produce both blackbody-like optical color temperature and Balmer jump ratios and track their evolution during the flare.


\section{New RHD Models}
The hot, blackbody-like (optical) phenomenon in dMe flares is notoriously difficult to reproduce in (1D) radiative-hydrodynamic (RHD) flare models using a solar-type electron "beam" heating mechanism \citep{Allred2006}.  Three beam parameters used as inputs to RHD models are the integrated energy flux density of the beam, which is denoted by an F\#, the power-law index ($\delta$) of the beam distribution, and the low-energy cutoff ($E_c$) of the beam's power law.  
By a solar-type beam, we mean beams with low-energy cutoffs of $\le 15-40$ keV, power-law indices of $\delta \sim 3-8$, and energy flux densities in the general range of F11 - F12 \citep{Milligan2014, Kuridze2015, Rubio2016, Warmuth2016, Kleint2016, Alaoui2017}, even though some solar events exhibit evidence of more extreme beam parameters \citep{Warmuth2009, Krucker2011, Thalmann2015, Ireland2013, Kerr2015}.  The TRACE white-light solar flare study of \citet{Fletcher2007} implies beam flux densities ranging from several times F11 to at least several times F12 (L. Fletcher, priv. communication) and lower-energy cutoff values of 17 keV if the white-light power is completely accounted for by the electron beam power.  However, spectral measurements of the Balmer jump and optical continuum are rare in solar flares.  A few spatially resolved observations in narrowband optical continuum bandpasses with Hinode/SOT (spanning a similar but slightly larger optical wavelength regime as two of the ULTRACAM filters) exhibit much lower color temperatures of $\le5000-6000$ K in the impulsive phase \citep{Watanabe2013, Kerr2014} than we calculate for the IF1 event on YZ CMi in Figure \ref{fig:yzcmi}. 

Extremely large electron beam energy flux densities ($10^{13}$ erg cm$^{-2}$ s$^{-1}$, ``F13'') have been used to study the dynamic atmospheric processes that can produce large continuum optical depth inferred from dMe observations \citep{Kowalski2015, Kowalski2016, Kowalski2017B}.   However, the current density of an F13 electron beam may produce plasma instabilities and a strong return current electric field \citep{Holman2012, Li2014}.  In these F13 RHD models calculated by the RADYN \citep{Carlsson1997} and RH \citep{Uitenbroek2001} codes, the hot blackbody-like spectra exhibits a relatively small Balmer jump in emission (like the observations) but lasts for only a short time.  
Furthermore, the instantaneous broadening of the hydrogen lines in the F13 models becomes extreme to a degree that has never been observed \citep{Kowalski2017B}.  The corona exceeds 50 MK, which only occurs in some dMe superflares \citep{Osten2010, Osten2016}, whereas both large and small dMe flares produce 10,000 K blackbody-like continuum radiation \citep{Kowalski2016}.  So how do these new high-time resolution inferences of color temperature in Figure \ref{fig:yzcmi} provide constraints and guidance for new RHD modeling of flares? 

Since stellar data do not have any direct spatial information of the flare, we consider two cases of spatial development that occur in solar flares. 

\begin{enumerate}
 
 \item The fast rise phase in Figure \ref{fig:yzcmi}(a) does not show any sub-structure on 1.1~s timescales.  If the impulsive phase source is just one or two very bright kernels, the monotonic ramp-up time of the beam flux density would be 20~s: $\Delta t_{\rm{fast-rise}} = \Delta t_{\rm{e-beam-burst}}$.  This is approximately the timescale range that has been found for beam-injection times in some solar flares \citep{Rubio2016}.  This interpretation implies evolution of the beam flux density but not an increase in flare area over which the beam energy is injected.  The peak beam flux density would have to be very large in order to explain the color temperature that forms at very early times in the rise phase.  RHD models with high beam flux densities are generally not run for longer than several seconds and are not available to test this scenario.  
 
 The projected surface areal coverage, $X \pi R_{*}^2$, of the blackbody-like emitting flare region is also fit to the IF1 event \citep[see][]{Hawley2003}).  Our fits to the blackbody-emitting area in the rise phase of IF show a monotonic increase, which suggests that a single, relatively long-duration heating event in a single flare loop is not a reasonable interpretation of the flare.  
The accuracy of blackbody radiation for determining flare areas should be tested against spatially resolved observations (as done for the Sun in \citet{Kretzschmar2011}) and verified against actual stellar atmosphere models.  Furthermore, a single color temperature parameterization implies a homogeneous source, which is likely not the case for unresolved stellar observations.  In conclusion, this modeling scenario is difficult to test given the uncertainties in the assumptions.


 \item   Our favored interpretation is that the fast rise phase of IF1 consists of many impulsively heated kernels for which the beam energy deposition burst is $\Delta t_{\rm{e-beam-burst}} << \Delta t_{\rm{fast-rise}}$.   This is called multithread modeling and was developed for spatially unresolved observations of solar flare X-rays from GOES \citep{Warren2006}.  Spatially resolved observations of solar flares support this scenario whereby sequentially heated flare loops (or ``threads'') form along a two-ribbon arcade, which spread apart during the peak and decay phases \citep{Kosovichev2001, Wang2007, Maurya2009, Qiu2010, Qiu2017}.  The lack of sub-structure in the fast rise of IF1 in Figure \ref{fig:yzcmi}(a)  suggests that the temporal separation between bursts along a hypothetical arcade/ribbons is $\le$1.1~s.  Future observations of solar flare optical kernels with the Daniel K. Inouye Solar Telescope will provide constraints on $\Delta t_{\rm{e-beam-burst}}$;  we have typically used impulsive beam heating durations of 2-3~s.  

A simple approach to multithread modeling is achieved by averaging an RHD model over its duration; this approach assumes that the same beam parameters are injected at all locations along the hypothetical stellar flare ribbons.  
We consider the burst-averaged, multithread beam models from Table 1 of \citet{Kowalski2017B}.  The ULTRACAM-like ratios (C4170/C6010) in this table are given for two F13 electron beam models: 1.6 and 1.8 which give values of $T_{\rm{FcolorR}} \sim 8000 - 9000$ K, respectively.  These multithread models actually give reasonable representations of many spectral properties in the early rise phase of the IF3 event from \citet{Kowalski2013} \citep{Kowalski2017B}.  

Though these F13 models attain an optical color temperature of $T\gtrsim 10,000$ K and a Balmer jump ratio of $\sim2$ at certain times in their evolution, the persistence of these continuum properties is not long enough (or quick enough to develop) in the burst-averaged model spectra in order to be consistent with the data of the fast rise phase of the IF1 event in Figure \ref{fig:yzcmi}. 
Models of chromospheric condensation evolution show they take some time ($t \ge 1.6$~s for very high beam fluxes) to increase in density and decrease in temperature from $T>40,000$ K to $T<13,000$ K such that they become optically thick at Balmer continuum wavelengths in the $U$-band.  The evolution of the optical color temperature in these F13 models changes rapidly from 7500 K (in the rise phase; at 1.2~s) to $T\gtrsim10,000$ K (at the peak; at 2-2.3~s) to 5500 K (during the fast decay; at 4~s).  This evolution of the color temperature is not seen at the 1.1~s cadence of the ULTRACAM flare data in Figure \ref{fig:yzcmi}(b).  The evolution of the Balmer jump ratio and color temperature are qualitatively reproduced in previous RHD models with F13 beams, but the changes occur on much shorter timescales than the color evolution from the fast rise to the fast decay that is observed in IF1 in Figure \ref{fig:yzcmi_colors}.  Perhaps, improving the input physics (e.g., opacities) of these F13 models may help bring their timescales closer to the observations.  

The evolution of flare colors provides guidance for modeling the actual time-evolution of flare light curves and colors. 
The ULTRACAM data of IF1 on YZ CMi motivate a grid of new radiative-hydrodynamic dMe flare models that produce hot blackbody-like radiation over a larger fraction of the evolution of an impulsive heating burst, particularly within the first two seconds.  Then, average burst (multithread) model flare spectra would have similar properties, like an optical color temperature $T\gtrsim10,000$ K, to many of the instantaneous model snapshots. The Cool Stars 20 poster contribution from Loyd et al.\ presented a color temperature (in the FUV) of $T \sim 15,500$ K during a superflare on a young dMe star (\emph{ApJ}, in press);  this further challenges the models to reproduce the extreme regimes of heating at large column mass in a multithread analysis.  

 Our new grid of radiative-hydrodoynamic flare models employs electron beams with high, low-energy cutoff values.  These beams significantly heat the atmosphere at large column mass.  In a future work, the results from this model grid will be discussed and compared to the observations and data analysis presented in these proceedings.  

\end{enumerate}

\section{Summary \& Conclusions}
The  $T=10^4$ K blackbody-like radiation has only been robustly constrained (with optical spectra or narrowband continuum ratios) to be energetically dominant during the impulsive phase of some impulsive-type flare (``IF'') events on very magnetically active mid-M dwarfs.   However, some dMe flares do not exhibit spectral evidence for such a large color temperature through the NUV, $U$-band, and optical (Kowalski et al. 2018, submitted to \emph{ApJ}). We present new measurements at 15x higher time resolution and 10x higher signal-to-noise than ever before during the fast rise phase of a large IF-type event that exhibits an optical continuum color temperature of $T\sim10^4$ K .  These observations provide new, compelling evidence for this phenomenon in dMe flares.

These data constrain the timescales of the formation of 10,000 K blackbody-like radiation in new radiative-hydrodynamic models.   The optical continuum is bluest (apparently hottest) in the fast rise phase, which also provides guidance for future photon-counting (``time-tagged'') observations from space that seek to characterize this continuum radiation at shorter wavelengths.


\begin{thebibliography}{50}
\providecommand{\natexlab}[1]{#1}

\bibitem[\protect\astroncite{{Alaoui} \& {Holman}}{2017}]{Alaoui2017}
{Alaoui}, M. \& {Holman}, G.~D. 2017, \apj, 851, 78.

\bibitem[\protect\astroncite{{Allred} \emph{et~al.}}{2006}]{Allred2006}
{Allred}, J.~C., {Hawley}, S.~L., {Abbett}, W.~P., \& {Carlsson}, M. 2006,
  \apj, 644, 484.

\bibitem[\protect\astroncite{{Ayres}}{2015}]{Ayres2015B}
{Ayres}, T.~R. 2015, \aj, 149, 58.

\bibitem[\protect\astroncite{{Carlsson} \& {Stein}}{1997}]{Carlsson1997}
{Carlsson}, M. \& {Stein}, R.~F. 1997, \apj, 481, 500.

\bibitem[\protect\astroncite{{Davenport} \emph{et~al.}}{2012}]{Davenport2012}
{Davenport}, J.~R.~A., {Becker}, A.~C., {Kowalski}, A.~F., {Hawley}, S.~L.,
  {Schmidt}, S.~J., \emph{et~al.} 2012, \apj, 748, 58.

\bibitem[\protect\astroncite{{Dhillon} \emph{et~al.}}{2007}]{Dhillon2007}
{Dhillon}, V.~S., {Marsh}, T.~R., {Stevenson}, M.~J., {Atkinson}, D.~C.,
  {Kerry}, P., \emph{et~al.} 2007, \mnras, 378, 825.

\bibitem[\protect\astroncite{{Fletcher} \emph{et~al.}}{2007}]{Fletcher2007}
{Fletcher}, L., {Hannah}, I.~G., {Hudson}, H.~S., \& {Metcalf}, T.~R. 2007,
  \apj, 656, 1187.

\bibitem[\protect\astroncite{{Fuhrmeister}
  \emph{et~al.}}{2008}]{Fuhrmeister2008}
{Fuhrmeister}, B., {Liefke}, C., {Schmitt}, J.~H.~M.~M., \& {Reiners}, A. 2008,
  \aap, 487, 293.

\bibitem[\protect\astroncite{{Hawley} \emph{et~al.}}{2003}]{Hawley2003}
{Hawley}, S.~L., {Allred}, J.~C., {Johns-Krull}, C.~M., {Fisher}, G.~H.,
  {Abbett}, W.~P., \emph{et~al.} 2003, \apj, 597, 535.

\bibitem[\protect\astroncite{{Hawley} \& {Pettersen}}{1991}]{HP91}
{Hawley}, S.~L. \& {Pettersen}, B.~R. 1991, \apj, 378, 725.

\bibitem[\protect\astroncite{{Holman}}{2012}]{Holman2012}
{Holman}, G.~D. 2012, \apj, 745, 52.

\bibitem[\protect\astroncite{{Howard} \emph{et~al.}}{2018}]{Howard2018}
{Howard}, W.~S., {Tilley}, M.~A., {Corbett}, H., {Youngblood}, A., {Loyd},
  R.~O.~P., \emph{et~al.} 2018, \apjl, 860, L30.

\bibitem[\protect\astroncite{{Ireland} \emph{et~al.}}{2013}]{Ireland2013}
{Ireland}, J., {Tolbert}, A.~K., {Schwartz}, R.~A., {Holman}, G.~D., \&
  {Dennis}, B.~R. 2013, \apj, 769, 89.

\bibitem[\protect\astroncite{{Kerr} \& {Fletcher}}{2014}]{Kerr2014}
{Kerr}, G.~S. \& {Fletcher}, L. 2014, \apj, 783, 98.

\bibitem[\protect\astroncite{{Kerr} \emph{et~al.}}{2015}]{Kerr2015}
{Kerr}, G.~S., {Sim{\~o}es}, P.~J.~A., {Qiu}, J., \& {Fletcher}, L. 2015, \aap,
  582, A50.

\bibitem[\protect\astroncite{{Kleint} \emph{et~al.}}{2016}]{Kleint2016}
{Kleint}, L., {Heinzel}, P., {Judge}, P., \& {Krucker}, S. 2016, \apj, 816, 88.

\bibitem[\protect\astroncite{{Kosovichev} \& {Zharkova}}{2001}]{Kosovichev2001}
{Kosovichev}, A.~G. \& {Zharkova}, V.~V. 2001, \apjl, 550, L105.

\bibitem[\protect\astroncite{{Kowalski}}{2015}]{Kowalski2015IAU}
{Kowalski}, A.~F. 2015, IAU General Assembly, 22, 2257997.

\bibitem[\protect\astroncite{{Kowalski} \emph{et~al.}}{2017}]{Kowalski2017B}
{Kowalski}, A.~F., {Allred}, J.~C., {Uitenbroek}, H., {Tremblay}, P.-E.,
  {Brown}, S., \emph{et~al.} 2017, \apj, 837, 125.

\bibitem[\protect\astroncite{{Kowalski} \emph{et~al.}}{2015}]{Kowalski2015}
{Kowalski}, A.~F., {Hawley}, S.~L., {Carlsson}, M., {Allred}, J.~C.,
  {Uitenbroek}, H., \emph{et~al.} 2015, \solphys, 290, 3487.

\bibitem[\protect\astroncite{{Kowalski} \emph{et~al.}}{2010}]{Kowalski2010}
{Kowalski}, A.~F., {Hawley}, S.~L., {Holtzman}, J.~A., {Wisniewski}, J.~P., \&
  {Hilton}, E.~J. 2010, \apjl, 714, L98.

\bibitem[\protect\astroncite{{Kowalski} \emph{et~al.}}{2013}]{Kowalski2013}
{Kowalski}, A.~F., {Hawley}, S.~L., {Wisniewski}, J.~P., {Osten}, R.~A.,
  {Hilton}, E.~J., \emph{et~al.} 2013, \apjs, 207, 15.

\bibitem[\protect\astroncite{{Kowalski} \emph{et~al.}}{2016}]{Kowalski2016}
{Kowalski}, A.~F., {Mathioudakis}, M., {Hawley}, S.~L., {Wisniewski}, J.~P.,
  {Dhillon}, V.~S., \emph{et~al.} 2016, \apj, 820, 95.

\bibitem[\protect\astroncite{{Kretzschmar}}{2011}]{Kretzschmar2011}
{Kretzschmar}, M. 2011, \aap, 530, A84.

\bibitem[\protect\astroncite{{Krucker} \emph{et~al.}}{2011}]{Krucker2011}
{Krucker}, S., {Hudson}, H.~S., {Jeffrey}, N.~L.~S., {Battaglia}, M., {Kontar},
  E.~P., \emph{et~al.} 2011, \apj, 739, 96.

\bibitem[\protect\astroncite{{Kunkel}}{1970}]{Kunkel1970}
{Kunkel}, W.~E. 1970, \apj, 161, 503.

\bibitem[\protect\astroncite{{Kuridze} \emph{et~al.}}{2015}]{Kuridze2015}
{Kuridze}, D., {Mathioudakis}, M., {Sim{\~o}es}, P.~J.~A., {Rouppe van der
  Voort}, L., {Carlsson}, M., \emph{et~al.} 2015, \apj, 813, 125.

\bibitem[\protect\astroncite{{Li} \emph{et~al.}}{2014}]{Li2014}
{Li}, T.~C., {Drake}, J.~F., \& {Swisdak}, M. 2014, \apj, 793, 7.

\bibitem[\protect\astroncite{{Loyd} \emph{et~al.}}{2018}]{Loyd2018A}
{Loyd}, R.~O.~P., {France}, K., {Youngblood}, A., {Schneider}, C., {Brown}, A.,
  \emph{et~al.} 2018, ArXiv e-prints.

\bibitem[\protect\astroncite{{Maurya} \& {Ambastha}}{2009}]{Maurya2009}
{Maurya}, R.~A. \& {Ambastha}, A. 2009, \solphys, 258, 31.

\bibitem[\protect\astroncite{{Milligan} \emph{et~al.}}{2014}]{Milligan2014}
{Milligan}, R.~O., {Kerr}, G.~S., {Dennis}, B.~R., {Hudson}, H.~S., {Fletcher},
  L., \emph{et~al.} 2014, \apj, 793, 70.

\bibitem[\protect\astroncite{{Mitra-Kraev}
  \emph{et~al.}}{2005}]{MitraKraev2005}
{Mitra-Kraev}, U., {Harra}, L.~K., {G{\"u}del}, M., {Audard}, M.,
  {Branduardi-Raymont}, G., \emph{et~al.} 2005, \aap, 431, 679.

\bibitem[\protect\astroncite{{Osten} \emph{et~al.}}{2010}]{Osten2010}
{Osten}, R.~A., {Godet}, O., {Drake}, S., {Tueller}, J., {Cummings}, J.,
  \emph{et~al.} 2010, \apj, 721, 785.

\bibitem[\protect\astroncite{{Osten} \emph{et~al.}}{2016}]{Osten2016}
{Osten}, R.~A., {Kowalski}, A., {Drake}, S.~A., {Krimm}, H., {Page}, K.,
  \emph{et~al.} 2016, \apj, 832, 174.

\bibitem[\protect\astroncite{{Qiu} \emph{et~al.}}{2010}]{Qiu2010}
{Qiu}, J., {Liu}, W., {Hill}, N., \& {Kazachenko}, M. 2010, \apj, 725, 319.

\bibitem[\protect\astroncite{{Qiu} \emph{et~al.}}{2017}]{Qiu2017}
{Qiu}, J., {Longcope}, D.~W., {Cassak}, P.~A., \& {Priest}, E.~R. 2017, \apj,
  838, 17.

\bibitem[\protect\astroncite{{Ranjan} \emph{et~al.}}{2017}]{Ranjan2017}
{Ranjan}, S., {Wordsworth}, R., \& {Sasselov}, D.~D. 2017, \apj, 843, 110.

\bibitem[\protect\astroncite{{Robrade} \& {Schmitt}}{2005}]{Robrade2005}
{Robrade}, J. \& {Schmitt}, J.~H.~M.~M. 2005, \aap, 435, 1073.

\bibitem[\protect\astroncite{{Rubio da Costa} \emph{et~al.}}{2016}]{Rubio2016}
{Rubio da Costa}, F., {Kleint}, L., {Petrosian}, V., {Liu}, W., \& {Allred},
  J.~C. 2016, \apj, 827, 38.

\bibitem[\protect\astroncite{{Segura} \emph{et~al.}}{2010}]{Segura2010}
{Segura}, A., {Walkowicz}, L.~M., {Meadows}, V., {Kasting}, J., \& {Hawley}, S.
  2010, Astrobiology, 10, 751.

\bibitem[\protect\astroncite{{Thalmann} \emph{et~al.}}{2015}]{Thalmann2015}
{Thalmann}, J.~K., {Su}, Y., {Temmer}, M., \& {Veronig}, A.~M. 2015, \apjl,
  801, L23.

\bibitem[\protect\astroncite{{Thomson} \emph{et~al.}}{2005}]{Thomson2005}
{Thomson}, N.~R., {Rodger}, C.~J., \& {Clilverd}, M.~A. 2005, Journal of
  Geophysical Research (Space Physics), 110, A06306.

\bibitem[\protect\astroncite{{Tilley} \emph{et~al.}}{2017}]{Tilley2017}
{Tilley}, M.~A., {Segura}, A., {Meadows}, V.~S., {Hawley}, S., \& {Davenport},
  J. 2017, ArXiv e-prints.

\bibitem[\protect\astroncite{{Uitenbroek}}{2001}]{Uitenbroek2001}
{Uitenbroek}, H. 2001, \apj, 557, 389.

\bibitem[\protect\astroncite{{Wang} \emph{et~al.}}{2007}]{Wang2007}
{Wang}, L., {Fang}, C., \& {Ming-DeDing} 2007, Chinese Journal of Astronomy and Astrophysics, 7, 721.

\bibitem[\protect\astroncite{{Warmuth} \emph{et~al.}}{2009}]{Warmuth2009}
{Warmuth}, A., {Holman}, G.~D., {Dennis}, B.~R., {Mann}, G., {Aurass}, H.,
  \emph{et~al.} 2009, \apj, 699, 917.

\bibitem[\protect\astroncite{{Warmuth} \& {Mann}}{2016}]{Warmuth2016}
{Warmuth}, A. \& {Mann}, G. 2016, \aap, 588, A115.

\bibitem[\protect\astroncite{{Warren}}{2006}]{Warren2006}
{Warren}, H.~P. 2006, \apj, 637, 522.

\bibitem[\protect\astroncite{{Watanabe} \emph{et~al.}}{2013}]{Watanabe2013}
{Watanabe}, K., {Shimizu}, T., {Masuda}, S., {Ichimoto}, K., \& {Ohno}, M.
  2013, \apj, 776, 123.

\bibitem[\protect\astroncite{{Zhilyaev} \emph{et~al.}}{2007}]{Zhilyaev2007}
{Zhilyaev}, B.~E., {Romanyuk}, Y.~O., {Svyatogorov}, O.~A., {Verlyuk}, I.~A.,
  {Kaminsky}, B., \emph{et~al.} 2007, \aap, 465, 235.

\end{thebibliography}

\end{document}